**ORIGINAL PAPER**

# Inhibition of Polyimide Photodegradation by Incorporation of Titanate Nanotubes into a Composite

Christian Harito[1,2,3] · Dmitry V. Bavykin[1] · Brian Yuliarto[2,3] · Hermawan K. Dipojono[3] · Frank C. Walsh[1]

**Abstract**

The effect of UV light exposure on the properties of hexafluoroisopropylidene-diphthalic anhydride–oxydianiline (6FDA–ODA) polyimide (PI) and polyimide–titanate nanotube (TiNT/PI) composites has been studied using Raman spectroscopy, optical microscopy, nanoindentation and TEM. The degree of polymer photodegradation was estimated by measuring the change in affinity to a positively charged dye (methylene blue, MB). The mechanism of photoassisted transformations in polyimides usually involves scission of polymer chains accompanied by appearance of active radicals, which undergo further rapid transformations to more stable phenol, amine, and carboxylic functional groups. The accumulation of these groups can increase the degree of adsorption of charged dyes in the photodegraded polymer. It was found that neat PI showed a significantly increased capacity to adsorb MB after irradiation with UV, reaching a plateau after 1 h. In contrast, TiNT/PI composite demonstrated a much slower rise in concentration of adsorbed MB even after 4 h of UV exposure. Raman spectra indicated cleavage of C=O and C–F bonds in PI while only the C–F bond was damaged in TiNT/PI. Shorter cracks ($\approx 40$ μm long) appeared in TiNT/PI composites whereas macro cracks (> 100 μm) were visible in neat PI after 3 h of UV exposure. Brittleness was studied by comparing plasticity index which varied from 0 to 1 (0 corresponding to the most brittle material and 1 the most ductile one). Plasticity index reduced by 51% and 2% for PI and TiNT/PI, respectively after 3 h UV irradiation, indicating that TiNT can protect underlying PI from further damage. The hardness of neat PI decreased whereas, for TiNT/PI, it increased under UV, suggesting crosslinking of broken polymer chains with nanotubes.

**Graphical Abstract**

Photodegradation of a titanate nanotubes/polyimide composite can lead to cross-linking of broken polymer chains by nano-structured material, resulting in increased hardness.

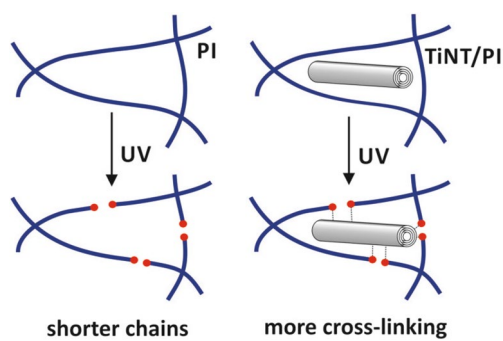

---

✉ Dmitry V. Bavykin
D.Bavykin@soton.ac.uk

Extended author information available on the last page of the article

# Introduction

Ultraviolet (UV) exposure is one of the most common environmental threats to polymer materials since it can lead to deterioration of properties and degradation. The functional groups of the polymer receive excess free energy after absorption of ultraviolet light forming unstable excited states [1]. Such unstable excited states may relax back to the ground state while producing heat or initiate breakdown of chemical bonds via formation of free radicals [2]. Formation of these free radicals may cause rearrangement of molecules, annihilation of small molecules, depolymerisation, crosslinking, or oxidation. Such chemical alterations may lead to discolouration of polymer and degradation of mechanical, thermal and electrical properties.

Ultraviolet degradation of polymer can be detected by both physical and chemical methods. Some physical changes such as roughness, micro cracks and surface pore cavities can be seen by atomic force, scanning electron, fluorescence or optical microscopy [3]. Contact angle measurement and wetting properties may indicate the effect of UV exposure on surface roughness and alteration of its chemical nature [4]. Melt rheology of the polymer is also changed due to scission or cross-linking of the polymer chain. Destructive methods such as tensile testing can elucidate the effect of UV on elasticity, strength, and modulus of bulk polymer [5]. Many modern physical tests (such as nanoindentation) are able to pinpoint various mechanical properties through thickness direction. Most of the reported nanoindentation studies of the UV stimulated degradation of the polymer materials have been focused on epoxy materials [6], so extension of this method to other polymer materials would be beneficial.

Chemical methods of characterisation can provide early detection of UV degradation of polymers. Electron spin resonance is able to detect free radicals within the polymer [7] and inhibition of UV degradation by incorporation of carbon nanotubes into polymers [8]. X-ray diffraction spectroscopy (XRD) and differential scanning calorimetry (DSC) can detect changes in the degree of crystallinity in crystalline or semi-crystalline polymers. Crystallinity is known to increase upon UV irradiation since amorphous parts of the polymer may degrade [9]. X-ray photoelectron spectroscopy (XPS) can determine the difference in binding energy between atoms on the polymer surface within 5–10 nm in depth. The method can also be coupled with time-of-flight secondary ion mass spectrometry (TOF-SIMS) providing imaging based on mass spectral data up to 1–2 nm [10]. Electrochemical impedance spectroscopy can trace biodegradation of polymers stimulated by microorganisms [11]. Chromatography can detect volatile organic compounds that may be released from the polymer bulk during UV irradiation. Oxidised polymer may also emit light when heated in an inert atmosphere, giving rise to chemiluminescence. Such luminescence might indicate the presence of carbonyl compounds and hydroperoxides in the polymer [12, 13]. Infrared spectroscopy and Raman spectroscopy is routinely employed to detect changes in the chemical bonding in the polymer matrix. Carbonyl index, which is the ratio of carbonyl bond intensity to the reference bond not affected by UV, is often used as an indication of UV degradation in the polymer [14, 15], despite anxieties that such an indicator is not fully representative due to variations in the mechanism of degradation. An integrated combination of characterisation methods may be the best way to study the UV degradation of polymers.

In this paper, our original method based on methylene blue (MB) sorption on moieties created by PI photodegradation [16] is adapted to estimate the degree of polymer photodegradation. Polyimide film is widely used in technological uses ranging from space applications and electronics, in which exposure to UV light can be the subject of photostimulated degradation, usually occurring via scission of polymer chains and formation of active radicals, relaxation of which can lead to formation of phenol, amine, isocyanic and carboxylic functional groups [17]. Such groups can become sorption centres for cationic methylene blue dye. The changes in methylene blue concentration can be easily measured by UV–Vis spectroscopy and reflect the changed in sorption capacity of degrading polyimide.

For inhibition of photodegradation, wide bandgap semiconductors, such as $TiO_2$ [18] or titanate nanotubes (TiNT) can be used as UV light absorber or screener. Depending on manufacturing technique, titanium oxide may have either strong or weak photocatalytic properties in reaction of non-selective oxidation of organic molecules, usually accompanied by photogeneration of OH radicals. Various crystallographic forms of $TiO_2$ can show different photocatalytic activity, with anatase typically displaying a highest response whereas rutile is less likely to be active and is more frequently used as a UV screener for poly(L-lactide). The TiNT used in this work are characterised by a trititanic acid monoclinic crystal structure ($H_2Ti_3O_7$) [19] and usually show poor photocatalytic activity for oxidation of organic molecules, probably due to sodium ion impurities, which introduce centres of recombination for photogenerated electrons and holes [20]. The incorporation of the TiNT into polymer matrixed can significantly improve the various properties of the composites [21]. The effectiveness of TiNT as a screener was determined by comparing PI and TiNT/PI composite degradation rates under UV light using Raman spectroscopy, nanoindentation, transmission electron microscopy (TEM) and optical microscopy. The increase in the number of broken PI chains during UV assisted photodegradation was estimated by measurements of concentration of MB sorption centres within a PI matrix. The unusual increase in hardness of TiNT/PI composite after UV exposure provides an important insight to the mechanism of cross-linking of broken polymer chains with nanostructured titanates.

# Experimental

## Preparation of Titanate Nanotubes

Our previous work [22] was used as a methodology for TiNT synthesis. As a typical example, 20 g of $TiO_2$ (Degussa P25, Aeroxide) was refluxed with 150 $cm^3$ of aqueous 10 mol $dm^{-3}$ solution of mixed KOH:NaOH (Sigma-Aldrich) with a molar ratio of 1:25 in a PFA (perfluoroalkoxy polymer) round-bottom flask (Bohlender GmbH) for 4 days at 106 °C and atmospheric pressure. The resulting powdered sodium titanate $Na_2Ti_3O_7$ was separated from alkali solution by filtration, followed by washing with distilled water until the solution reached pH 7. The protonated titanate $H_2Ti_3O_7$ was formed by washing sodium titanate with an excess of 0.1 mol $dm^{-3}$ HCl (Sigma-Aldrich) for more than 30 min until a stable pH of 2 was reached, then distilled water washing to pH 5. Finally, 0.2 g of $H_2Ti_3O_7$ was sonicated for 9 h with 50 $cm^3$ of 0.075 mol $dm^{-3}$ aqueous tetramethylammonium hydroxide (TMAOH). The solution was kept for 4 days to separate isolated nanotubes (at the top) from non-dispersed agglomerates (at the bottom). The stable colloidal solution on the top was used to prepare the composites.

## Preparation of 6FDA–ODA Polyimide

A partially fluorinated PI was synthesised from 4,4′-(hexafluoroisopropylidene)diphthalic anhydride (6FDA) and 4,4′-Oxydianiline (ODA) as monomers using a two-step (polyamic acid formation followed by its imidization) method [23]. As a typical example, 1.553 g of ODA was dissolved in 20 g of N,N-dimethylformamide (DMF) in a 250 $cm^3$ round bottom flask. 3.447 g (equimolar) of 6FDA was slowly added to the solution and stirred for 2 days at $22.5 \pm 2.5$ °C yielding 20 wt% of 6FDA–ODA polyamic acid solution. The polymerisation reaction was shielded with nitrogen gas. A mixture of pyridine and acetic anhydride (3.5:4 volume ratio) was used for chemical imidization. 15 $cm^3$ of imidization solution was poured slowly into polyamic acid solution and stirred for 1 day. The final solution was slowly precipitated in methanol, filtered, and washed with methanol. The resulting yellowish powder was dried in vacuum at $22.5 \pm 2.5$ °C for 1 day. For pure polyimide film, 7.764 g of PI powder was dissolved in 40 $cm^3$ dimethyl sulfoxide (DMSO) by stirring for 2 weeks. 140 μl of the solution was taken and casted on top of the glass followed by drying at 120 °C for 1 h.

## Preparation of 6FDA-ODA/Titanate Nanotubes Nanocomposite

The composite film was prepared by evaporation of mixed solutions of dissolved PI and suspended TiNT. 7.376 g of 6FDA–ODA PI powder was stirred with 38 ml DMSO at room temperature over 2 weeks until it fully dissolved. An aqueous colloidal solution of TiNT was concentrated in vacuum rotary evaporator at 70 °C to reduce the volume to 2 $cm^3$ containing 0.0745 g TiNT (1 wt% TiNT final solid composite). 2 $cm^3$ of concentrated TiNT was slowly added to 38 $cm^3$ of PI solution and stirred for 1 week to form a homogeneous mixture. 140 μl of final solution was taken after stirring and casted on the glass followed by drying in a furnace at 120 °C for 1 h.

## Characterisation

Raman spectra were recorded using a Raman spectroscopy confocal microscope (Renishaw, RM 2000) with excitation wavelength at 632.8 nm. The exposure time was 10 s with a 1% intensity of laser radiation. To quantify the spectral changes of PI exposed to UV radiation, the peak area $A_g$ related to unstable functional group such as C=O or C–F was referenced against the peak area, $A_{ph}$ of a stable functional group, which in the case of PI is the aromatic C=C bond at a wavenumber of $\approx 1618$ $cm^{-1}$ using an equation suggested elsewhere [24]:

$$\Delta A = \frac{A_g/A_{ph} - A_g^0/A_{ph}^0}{n \times A_g^0/A_{ph}^0} \quad (1)$$

where $A_g^0$ and $A_g$ are initial (prior to UV exposure) and current (after UV exposure) peak areas of specific functional group affected by photodegradation respectively, $A_{ph}^0$ and $A_{ph}$ are initial and current peak areas the areas from aromatic C=C bond respectively and $n$ is the number of specific functional groups in one repeating unit of 6FDA–ODA polyimide. (e.g., 4 for C=O, 6 for C–F). The relative area change ($\Delta A$) evaluated from Eq. (1) shows the number of degrading specific functional groups under UV irradiation. A more negative $\Delta A$ value indicates that more groups in the polymer chain have been cleaved.

Transmission electron microscopy (TEM) (JEOL-3010) was employed to study the dispersion of TiNT in the PI. Dissolved solution of PI in DMSO was mixed with an aqueous suspension of TiNT and the mixture was dropped on top of a copper grid with a perforated carbon film followed by spin coating of the solution at 4300 rpm for 1 min. The obtained thin film of composite on the copper grid was transparent to an electron beam in TEM.

The optical photograph of 6FDA–ODA PI and TiNT/PI film was taken by a 20.1 megapixels camera and Celestron™ LCD Digital Microscope II.

Nanoindentation was performed on a Nanotest Platform 3 nanoindenter (Micro Materials Ltd, UK) to obtain reduced elastic modulus and hardness. To prepare the samples, a small cut of the film (around 0.5 cm × 0.5 cm) was glued by epoxy glue to the soda-lime glass with the smoother surface (cast side) on top to be measured. The thickness of the polymer film (measured by screw gauge micrometre) was 100 μm. The thickness had varied from sample to sample in range from 90 to 110 μm. The other side of the glass was fixed with acrylic glue to a cylinder holder. The holder, complete with glass and samples on top, was put inside the nanoindenter. A Berkovich (3-side pyramidal) diamond tip was used for nanoindentation measurements. A constant loading and unloading rate of 20 mN s⁻¹ were applied with a 30 s holding time after reaching the maximum depth (ca 1 μm). 0.5 and 1 mN applied loading ranges were set. At least 20 indentations were performed at each applied load, for all samples.

### UV Photodegradation Studies of Polymer Composites and Estimation of Polymer Chain Scission Rate by Sorption of MB

All film samples of PI and TiNT/PI composites were illuminated with a parallel beam of unfiltered light from a Luxtel light source equipped with 300 W CeraLux CL300BUV-10F xenon arc lamp. The distance between the sample and the UV lamp was 13 cm. Samples were irradiated for 1, 2, 3, and 4 h at $22 \pm 3$ °C inside a ventilated fume hood. The incident power density of light at the point of sample holder was determined using UV strips (Alpha-Cure) to be approximately 1.5 W cm⁻².

In order to estimate the amount of broken polymer chains after exposure of PI to UV light, the capacity of the film to adsorb MB cations was determined using the following procedure. A film sample, 1 cm × 1 cm in size, was weighed and immersed in 5 cm³ of $1 \times 10^{-6}$ mol dm⁻³ methylene blue (MB) aqueous solution for 24 h at $22 \pm 2$ °C. The concentration of MB before ($C_0$) and after ($C$) sorption was determined using a UV–Vis spectrometer Neosys-2000 (Scinco). A molar extinction coefficient of 708 dm³ mol⁻¹ cm⁻¹ was used at a wavelength of 667 nm. Each sample was measured twice and the average was taken to determine the concentration. The amount of methylene blue ($a^{PI}$) absorbed by a neat 6FDA–ODA PI sample was calculated using:

$$a^{PI} = \frac{(C_0 - C)V}{m} \tag{2}$$

where $V$ is the volume of methylene blue solution and $m$ is the mass of the film sample. When PI composite containing TiNT was used, the amount of MB sorbed by the sample ($a^{PI/TiNT}$) has contributions, from both pure polymer ($a^{PI}$) and titanate nanotubes ($a^{TiNT}$):

$$a^{PI/TiNT} = a^{PI} + a^{TiNT} \tag{3}$$

The value $a^{TiNT}$ depends mostly on the surface charge and specific area of TiNT; it is not affected by UV radiation [25] and can be determined by comparisons between $a_0^{PI/TiNT}$ and $a_0^{PI}$ under dark conditions (prior to UV exposure) for composite and neat polymer, respectively.

$$a^{TiNT} = a_0^{TiNT} = a_0^{PI/TiNT} - a_0^{PI} \tag{4}$$

The amount of MB sorbed by photodegrading polymer ($a^{PI}$) for neat and composite polymer material after their UV exposure was calculated by combining Eqs. (2)–(4).

## Results and Discussion

### Measurement of the Degree of Photodegradation

Mechanisms of photoassisted degradation of polymers are usually complex and involve many pathways for the chemical reactions due to the high activity of photogenerated radicals, which subsequently unselectively attack polymer chain resulting in multiple products of decomposition [26]. Nevertheless, the common initial reaction during photodegradation is scission of polymer chains forming at least two chain ends. These ends can either rapidly react with other polymer chains leading to cross-linking or participate in series of transformations resulting in formation of less active ends of the chain. Such ends of the chain usually have specific functional groups which are normally absent in original (not degraded polymer).

In case of polyimide, UV irradiation cause generation of various active radicals [27], which undergo further transformation leading to formation of specific functional groups at the ends of broken chains. For example, photodegradation of PMDA–ODA polyimide (produced by polycondensation of pyromellitic dianhydride with oxydianiline) produces phenyl isocyanate [28], phenol, amine groups [17], and volatile gases (e.g., CO, $CO_2$) [28]. XPS studies have revealed that polymer chain scission is most likely initially occurs in C–N bonds in amide group (N–C=O) as well as in C–O bonds in either group (C–O–C) [17]. In contrast, inner carbon–carbon bonds in conjugated aromatic structures are less likely to be broken during photodegradation. Figure 1 indicates the chemical bonds by red dotted lines which are more likely to be cleaved during photodegradation of 6FDA–ODA

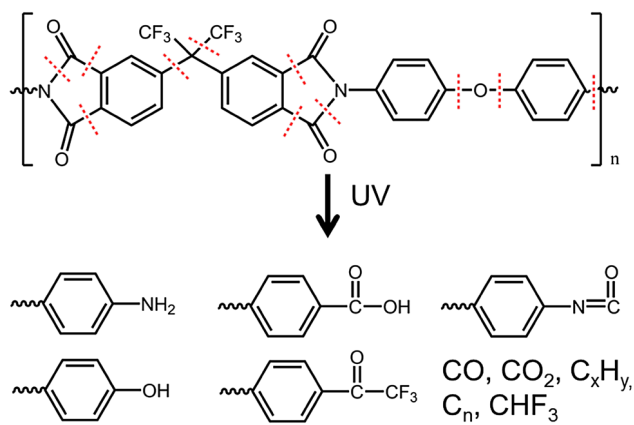

**Fig. 1** Possible positions (indicated by red dashed lines) of 6FDA–ODA polyimide chain scission during UV assisted photodegradation leading to formation following functional groups at the chain ends (Color figure online)

polymer chain. The weakest bonds were analogous to PMDA–ODA polyimide. Considering high dissociation energy of C=O (748 kJ mol$^{-1}$) and C–F (487 kJ mol$^{-1}$) bonds [29, 30], they are less likely to be broken, compared to neighbouring C–C (349 kJ mol$^{-1}$), C–N (307 kJ mol$^{-1}$), C–O (360 kJ mol$^{-1}$), and N–CO (361 kJ mol$^{-1}$). Therefore, photodegradation of 6FDA-ODA polyimide results in scission its polymer chains ends of which typically have phenol, amine, isocyanate, and carboxylic acid functional groups as well as forming volatile low molecular weight substances such as CO, CO$_2$, and CHF$_3$ (see Fig. 1).

It is important to note that the appearance of polar functional groups at the ends of the broken chains of photodegraded polymer may increase its affinity to polar dye molecules. As a result, the degree of photodegradation can be monitored by measuring the polymer's capacity to sorb charged dye molecules in the bulk matrix. We have found neat 6FDA–ODA to have a relatively low capacity to sorb methylene blue (MB) dye molecules from aqueous solutions, whereas samples irradiated by UV are characterised by a much higher capacity (see Fig. 2a (right) and b (right)). This phenomenon is consistent with the above mechanism of PI photodecomposition and can be used to estimate the quantity of such broken chains in the polymer, assuming that it is proportional to the quantity of generated adsorption sites for the MB. The concentration of these MB adsorption sites can be easily detected by measuring PI sorption capacity for MB molecules using Eq. (2) (see "Experimental" section).

The MB in aqueous solutions at neutral pH occurs usually in its cationic form and therefore it can strongly be attracted to negatively charged carboxylic or phenol groups at the ends of broken chains. In contrast, methyl orange (MO) at the same conditions occurs predominantly in anionic form. Staining both neat and UV irradiated PI with MO showed no sorption of dye into the polymer matrix, indicating the cationic nature of the functional groups and the broken chains of the polymer.

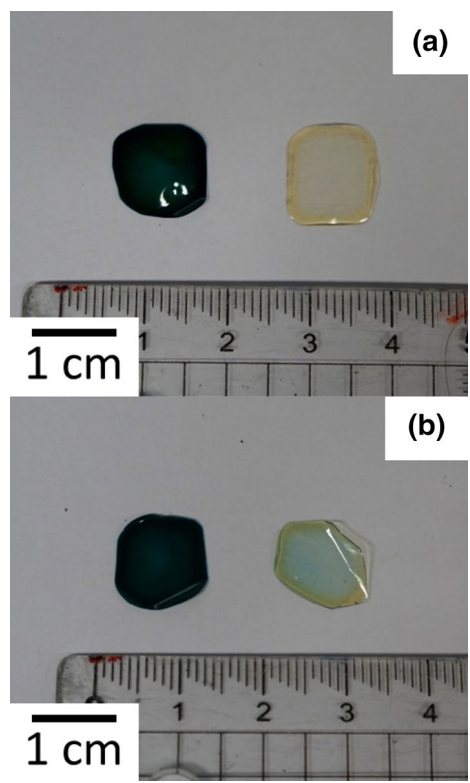

**Fig. 2** The photograph of neat and exposed to UV 6FDA–ODA polyimide films immersed for 24 h in aqueous methylene blue (10$^{-6}$ mol dm$^{-3}$). **a** non-irradiated TiNT/PI (left) and PI (right) samples; **b** TiNT/PI (left) and PI (right) films exposed for 4 h to UV

Figure 3 (black curve) shows accumulation of MB adsorption sites in pure PI during exposure of the polymer film to UV light. The kinetic curve is characterised by rapid four-fold increase in concentration of adsorption sites in polymer

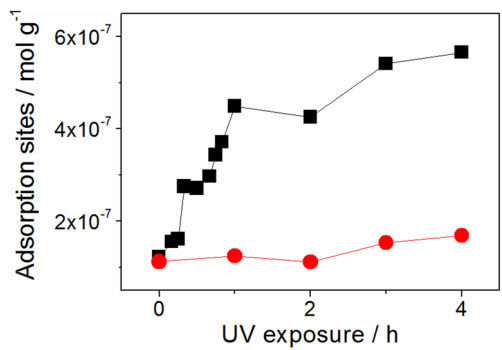

**Fig. 3** Growth in concentration of MB adsorption sites in neat PI (filled square) and TiNT/PI composite (filled circle) samples as a function of UV irradiation time (Color figure online)

within 1 h of UV exposure, followed by plateau after 4 h. Although there is no indication that photodegradation of polymer stops after 4 h of exposure, the presence of the plateau in the kinetic curve of accumulation of MB adsorption sites probably suggests appearance of additional pass for degradation of functional groups, which can sorb MB molecules. The contribution of that pass becomes significant only after long irradiation of polymer when sufficient quantity of these functional groups has been reached. In that case, the rate of generation of these functional groups becomes equal to the rate of their further degradation resulting in plateau. The initial linear part of the kinetic curve can be used for assessing the degree of photodegradation of the polyimide at early stages of decomposition as well as for investigation of the different factors which can promote (or inhibit) this process.

## Inhibition of Photodegradation by Addition of TiNT

We have fond that incorporation of TiNT into a PI matrix resulted in a significant increase in MB uptake in the mixed composite. Figure 2 (left) shows pictures of TiNT/PI film samples stained by MB. It is clear that compare to pure PI, the composite can adsorb much larger quantity of the dye molecules. This result was anticipated since, due to anionic nature of the surface, TiNTs are characterized by strong interactions with cationic types of dyes. In case of MB, a single nanotube can adsorb more than 1000 molecules of the dye from aqueous solutions [25]. Good adsorption properties of TiNT/PI composite indicate high permeability of the dye molecules through the PI membrane allowing MB to access the surface of TiNT. This observation also confirms that poor staining of pure PI and its subsequent increase after UV exposure (see Fig. 2a (right) and b (right)) was due to initial lack of MB adsorption centres, which concentration growth during photodegradation, rather than slow diffusion of the MB through the polymer matrix. One can speculate that, if the network of PI chains in the matrix is too dense for facile permeation of MB molecules though the polymer, the increase in MB uptake in neat PI after UV exposure could be attributed to an increase in the permeability accompanied by photodegradation of polymer and general loosening density of the chain network resulting in faster MB diffusion and its higher loading. However, such a hypothesis cannot explain a high MB uptake in unexposed to UV TiNT/PI composite, which demonstrates a good sorption capacity even after 24 h of staining. Moreover, analysis of TEM images of TiNT/PI composite in Fig. 4 confirms homogeneous distribution of nanotubes thought the bulk of polymer matrix indicating the depth of MB penetration through the initial polymer. It can be concluded that incorporation TiNT into PI does not significantly affect the adsorption properties of nanotubes since the polyimide will neither block its surface nor restrict the transport of dye.

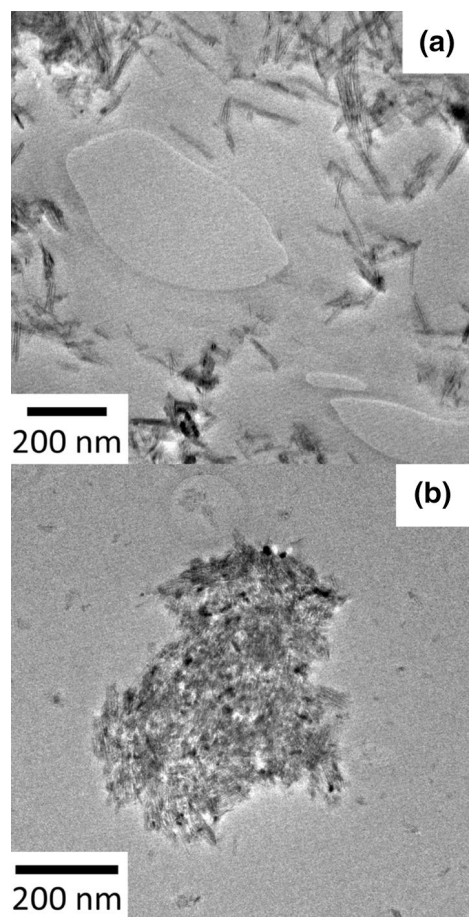

**Fig. 4** TEM images showing **a** distribution of TiNT in PI matrix; **b** occasional TiNT agglomerate

Since the adsorption of MB on TiNT has been previously well characterised, it is possible to take into account the contribution from MB molecules attached to the surface of TiNT and exclude that contribution in the mixed composite. Using a combination of Eqs. (2)–(4) (see "Experimental" section), the amount of MB attached to the polyimide only in the TiNT/PI composite was estimated. The red curve in Fig. 3 shows the dynamics of the MB adsorption centres growth as a function of UV exposure during photodegradation of PI reinforced with TiNT. In the presence of TiNT, the concentration of MB adsorption centres in the PI grow much more slowly compared to the case of pure polymer. Four hours irradiation of TiNT/PI composite increases the concentration of MB adsorption centres from $1.1 \times 10^{-7}$ to $1.7 \times 10^{-7}$ mol g$^{-1}$ whereas, in neat PI, the concentration grows from $1.2 \times 10^{-7}$ to $5.7 \times 10^{-7}$ mol g$^{-1}$. This corresponds to an almost three times lowering in the rate of generation of MB adsorption centres and consequently decrease in the rate of polymer chain scission after addition of nanotubes, which can be associated with inhibition of polymer photodegradation by TiNT.

The typical mechanism of the polymer protection against its photodegradation using TiO$_2$ based additives is considered to be mainly due to UV light blocking by added wide bandgap semiconductors which are transparent for visible light and opaque for short wavelength UV radiation. Although TiO$_2$ materials are well known for their high photocatalytic activity in reaction of oxidation of various organic substances [31], employment of TiO$_2$ as a polymer protector would require suppression of its activity. Inhibition of photocatalytic properties of TiO$_2$ is usually achieved by either blocking its surface or stimulation of recombination of photogenerated electrons and holes. TiNTs are usually produced by alkaline hydrothermal process and they contain high level of sodium ions impurities which are well known centres of electron hole recombination. As a results TiNT demonstrates poor photocatalytic activity in reaction of oxidation, which is useful for its use and polymer UV protector.

In order to understand the mechanism of TiNT assisted inhibition of photodegradation rate in polyimide materials, Raman spectroscopy was used to detect any intermediate moieties or determine which type of bonds are more likely to be cleaved under UV irradiations. Such studies of polymers photodegradation by monitoring characteristic chemical bonds vibrations using infrared or Raman spectroscopy were extensively used in past [32]. The carbonyl group (C=O) is the most common indicator of UV degradation [14, 15]. However, it has been also argued that carbonyl index may not be the best indicator of UV degradation of polypropylene since it takes a long time to see changes in this index making it unsuitable for early detection of UV degradation. Unlike polypropylene, fluorinated polyimide has several functional groups such as C–F and C–N including carbonyl groups that can be a good early indicator of UV damage in PI [24]. Raman spectra of 6FDA–ODA PI and TiNT/PI composite in Fig. 5, which show characteristic peaks from following functional groups C=C, C=O, C–N, C–F, were used to study photodegradation of PI and its composite.

The stretching vibrations ($\nu$) at wavenumbers of 1783 cm$^{-1}$, 1618 cm$^{-1}$, 1380 cm$^{-1}$, and 756 cm$^{-1}$ are attributed to C=O, C=C from phenyl, C–N, and C–F from CF$_3$, respectively. C–N bond in 6FDA–ODA polyimide overlapped with neighbouring peaks, making it difficult to calculate the area for photodegradation study. Therefore, C-N was excluded from the study. Phenyl C=C bonding in 6FDA–ODA polyimide is less susceptible to breakdown upon UV irradiation, rendering it useful as a reference peak. The change in absorbance of C=O and C–F was monitored for UV degradation. It is shown in Table 1 that the intensities of peaks for C–F and C=O bonds were decreased significantly after 3 h during UV irradiation of the pure PI sample. The relative area change $\Delta A$ for C=O bond was −0.079 while for C–F was −0.054, demonstrating that the C=O bond is more prone to breakdown by UV irradiation. This result agrees with a previous study photodegradation studies of PI [24].

In contrast, TiNT/PI composite film still retained C=O bonding after 3 h ($\Delta A = 0.009$) while the C–F bonds have cleaved but not in the same degree as in pure PI ($\Delta A = -0.035$). This indicates that non-calcined hydrothermally synthesised TiNT protect the most vulnerable bond (e.g., C=O) from UV attack and also reduce the damage of C–F bonds in 6FDA–ODA polyimide.

Despite the disappearance of reflections related to C=O and C–F bonds during photodegradation of PI and TiNT/PI composites, there are no new peaks in Raman spectra which could be related to polar functional groups at the ends of the broken polymer chains. Such a discrepancy with MB sorption results is probably due to the low sensitivity of Raman spectroscopy, rendering it unsuitable for detection of relatively low concentration of MB sorption centres at early stages of degradation. Nevertheless, both MB sorption experiments and Raman data indicate the protective role of TiNT, which inhibits photodegradation of the polymer chains.

Although the MB staining method is very useful for detection of polymer photodegradation at early stages, it relies on measurement of the MB sorbed within the polymer matrix rather than direct observation of broken polymer chains. We have found, however, that correlation between the quantity of broken polymer chains and the amount of sorbed MB is not always valid since MB within polymer matrix can undergoes slow decomposition by stable photogenerated reactive species. For example, it has been observed that UV irradiated PI samples stained with MB (same as in Fig. 2b right) bleach slowly at room temperature under dark conditions within 24 h. Such discoloration of the MB is probably caused by its oxidation by stable remained photogenerated radicals within polymer matrix [27, 33] slightly underestimating the concentration of broken polymer chains.

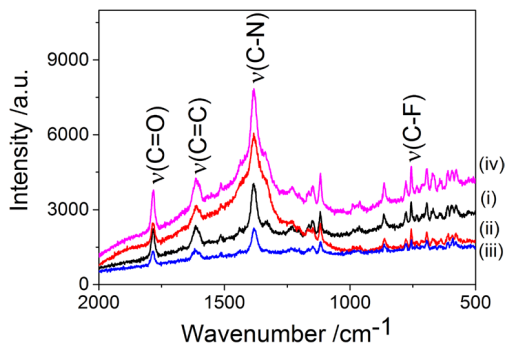

**Fig. 5** Raman spectra of PI and TiNT/PI (i, black) PI irradiated for 3 h; (ii, red) PI non-irradiated; (iii, blue) TiNT/PI irradiated for 3 h; (iv, purple) TiNT/PI non-irradiated (Color figure online)

**Table 1** The relative area change ($\Delta A$) for specific of C–F and C=O bonds calculated from Eq. (1) for PI and TiNT/PI composited after 3 h UV irradiation

| Group index | $\Delta A_{PI}$ | $\Delta A_{PI/TiNT}$ |
| --- | --- | --- |
| C=O index | −0.0787 | 0.0088 |
| C–F index | −0.0540 | −0.0350 |

## Effect of UV Irradiation on Physical Properties of the Polyimide Composites

The practical use of the polymer materials, in large extend, relies on their mechanical properties in various applications; therefore, the fist noticeable consequence of polymer degradation is alteration of these properties. The connection between the degree of the polymer photodegradation and extend of deterioration of its mechanical properties can be complex and dependant on the type of the polymer. Moreover, each polymer can have various bonds that produce different type of macro-radical depending on its dissociation energy [34]. Such a large number of possibilities create diverse physical changes such as reduced ductility, crosslinking [35], decolouration or yellowing [36], loss of transparency, reduction in toughness, cracking, and chalking due to erosion of the polymer surface, exposing filler or plasticisers. In this work, digital microscopy and nanoindentation techniques were used to monitor physical changes of irradiated PI and its nanocomposite with nanotubes TiNT/PI films.

Optical microscope images of top surface of the pure PI and TiNT/PI films exposed to UV are shown in Fig. 6. Neat initial polymer doesn't show any cracks prior UV irradiation (Fig. 6a) whereas, after 3 h of irradiation, both PI (Fig. 6c) and TiNT/PI composite (Fig. 6b) show visible micro-cracks of approximately 50–100 µm long. Pure PI samples, however, also have larger macro-cracks, > 100 µm long. Appearance of these photogenerated imperfections can be linked to the photodegradation of the polymer during which, various (such as C=O and C–F) covalent chemical bonds are cleaved reducing the average length of the polymer chain causing local stress and deformations within the polymer matrix. Inhibition of the PI photodegradation rate by addition of TiNT can reduce the rate of appearance of these microcracks. It is also possible that elongated ceramic nanostructures can hinder the propagation of the crack by alignment in the perpendicular direction to the crack inside stretched fibrils of the matrix [21], improving the general UV stability of the composite materials.

Addition of nanostructured fillers to polymers can greatly modify their properties and has been widely used in technological applications [37]. Distribution of TiNT filler within a PI matrix only marginally increases micro-hardness and Young's modulus (see Fig. 7 at 0 h) probably due to the low concentration of TiNT (1 wt%) in composite. We have previously shown that improvement of mechanical properties of the TiNT/PI composites is more evident at higher filler loadings [38].

The deterioration of the mechanical properties of PI and TiNT/PI composited under UV exposure can be directly

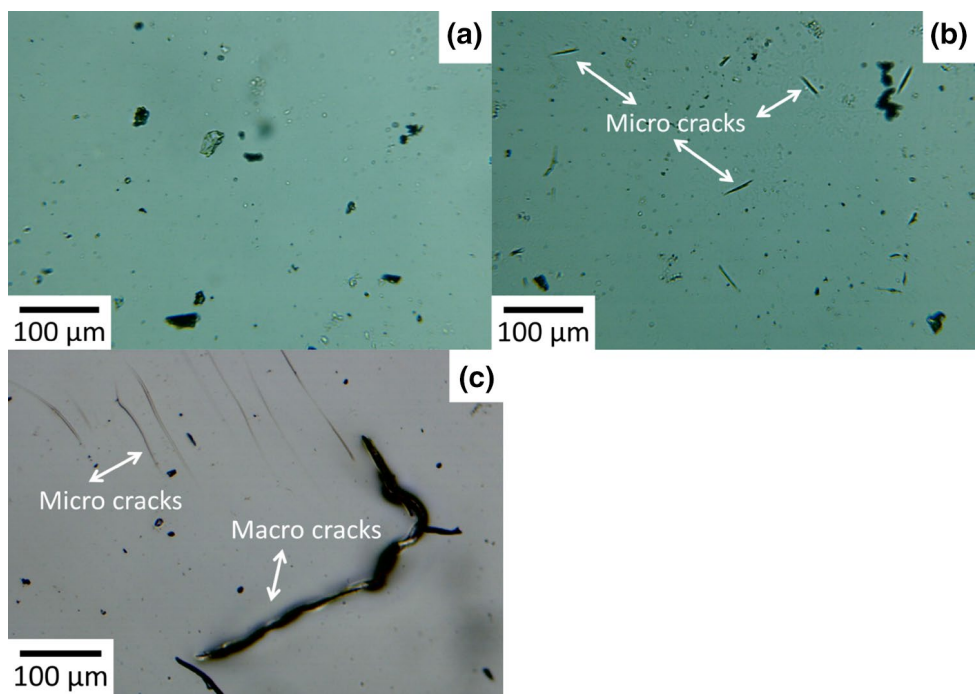

**Fig. 6** Digital optical microscopy images for **a** non-irradiated PI; **b** TiNT/PI after 3 h irradiation; **c** PI after 3 h UV exposure

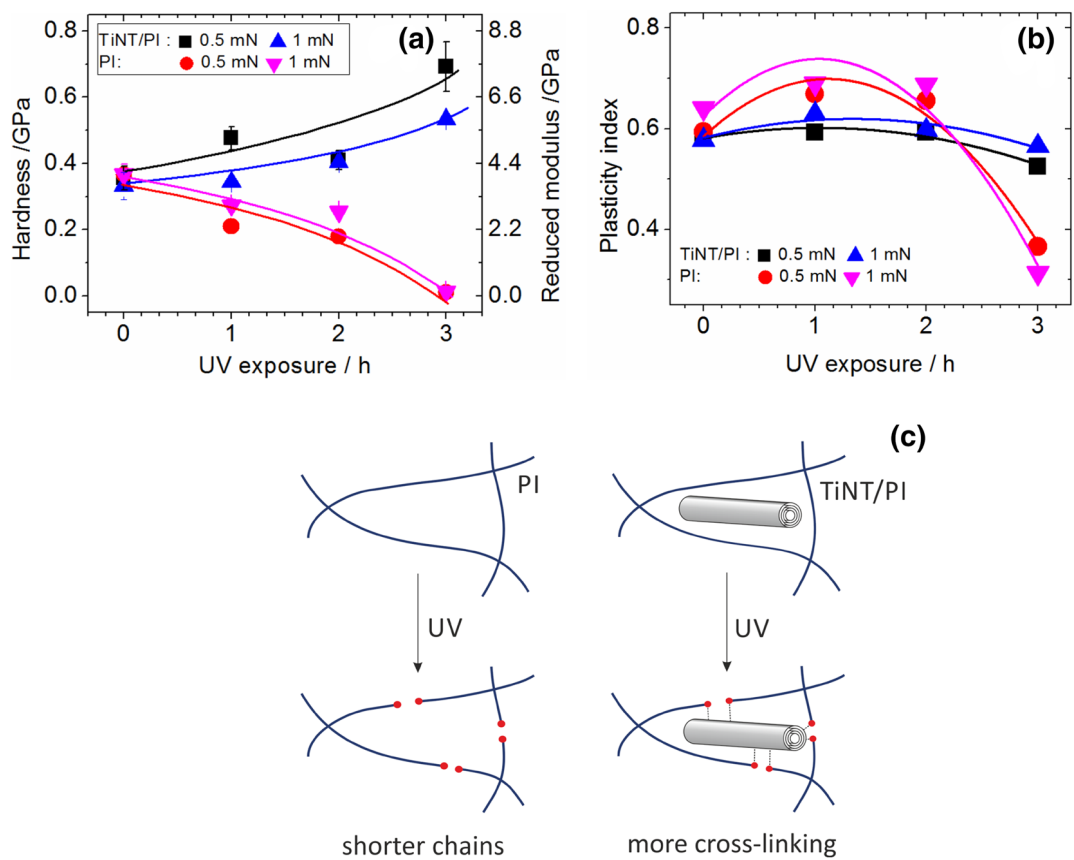

**Fig. 7** Nanoindentation studies of neat PI and TiNT/PI composite with 0.5 mN and 1 mN applied load; **a** hardness and reduced modulus, **b** plasticity index as a function of UV exposure time, **c** illustration of the crosslinking between TiNT and irradiated PI by additional bonds between broken chains and nanotubes

measured using nanoindentation technique. The dynamics of Young's modulus, micro-hardness, and plasticity index changes as a function of irradiation time for neat PI and TiNT/PI composite is shown in Fig. 7. The apparent tendency of the hardness and the modulus change for pure PI is contrasted with that for TiNT/PI nanocomposite. It was found that UV irradiation has increased the hardness and modulus of TiNT/PI composites while it has reduced that in neat PI polymer. The effect is more prominent after 3 h of exposure. Such difference in polymer behaviour after UV irradiation probably reflects the difference in mechanism of photodegradation of neat and TiNT filled poliimide.

Usually, the amorphous regions of polymer are more susceptible to UV degradation than crystalline. 6FDA–ODA polyimide is predominantly consisted of amorphous phase with low chain packing order due to the presence of bulky hexafluoroisopropylidene groups [39]. Chain cleavage on such amorphous PI reduces the average length of the polymer chain, which usually cause the decrease in polymer viscosity and reduction its hardness and modulus (see Fig. 7a). In case of added TiNT filler, however, the ends of broken chains may interact with dispersed in matrix nanotubes forming strong bonds with surface titanate groups. The exact nature of these possible bonds is still unknown. However, if several broken chains can be attached to a nanotube, it would provide additional crosslinking between several sliding polymer chains, where TiNT acts as a cross linker (see Fig. 7c). Crosslinking of filler and polymer matrix is known to strengthen interfacial bond between filler and matrix leading to higher tensile strength and impact resistance of the composite [40]. Also, crosslinking of broken polymer chains with elongated ceramic nanostructures is building the three dimensional network, which decrease the mobility of individual polymer chain, which is consistent with observed increase of the hardness and Young's modulus of the composite after UV exposure.

Brittleness may occur due to UV attack. Unlike hardness or modulus, plasticity index has strong correlation with brittleness. Plasticity index is a dimensionless property reflecting the ability of material to plastically deform before it breaks [41], which can be expressed by:

$$\text{Plasticity index} = W_p / W_p + W_e \qquad (5)$$

where $W_p$ and $W_e$ are the plastic work and elastic work done during nanoindentation, respectively. The value of plasticity index is varied from 0 to 1, where 0 for the most brittle material and 1 for the most ductile material. The plasticity index for PI and TiNT/PI is shown in Fig. 7b. For 0.5 mN applied load, the plasticity index of pure PI sample reduced significantly after 3 h of UV irradiation (38%) while only slight decreased observed for TiNT/PI (10%) compared to non-irradiated samples. The shielding effect of TiNT is more prominent at 1 mN applied load. Plasticity index reduced by 51% and 2% for PI and TiNT/PI, respectively after 3 h UV irradiation indicating that TiNT can protect PI matrix from further damage.

## Conclusions

The effect of UV exposure on 6FDA–ODA polyimide and polyimide–titanate nanotubes nanocomposite has been studied. In order to estimate the degree of PI photodegradation, the original method of staining polymer material with aqueous solution of methylene blue (MB) was developed. It was found that UV assisted photodegradation of PI is accompanied by scission of the polymer chain and generation of MB adsorption sites within the bulk of the polymer. The exact structure of these sites is unclear but comparative studies with methyl orange (MO) dye suggest their anionic nature. Addition of TiNT filler inhibited the rate of generation of these MB sorption sites decreasing photodegradation of PI. Raman spectroscopy also confirms the protecting role of TiNT additives by showing a reduced rate of disappearance of C=O and C–F groups in the TiNT/PI composite under UV irradiations.

The deterioration of mechanical properties and physical appearance of PI and TiNT/PI films was studied by nanoindentation and optical microscopy respectively. Photodegradation of neat PI results in appearance of macro-cracks with length more than 100 μm, whereas composite shows only micro-cracks with typical length in range of 50–100 μm after 3 h of UV exposure. It was also found that hardness and modulus is increased in TiNT/PI composite but decreased in neat PI under UV irradiations suggesting fundamental difference in the mechanism of degradation of the polymer. UV light stimulated reactions in neat PI usually leads to scission of the polymer chain and reduction in its average length whereas, in the presence of titanate nanotubes, such broken polymer chains have tendency to recombine with the elongated TiNT crosslinking the PI matrix and consequently increasing the composite hardness and Young's modulus. Brittleness was studied by comparing plasticity index which varied from 0 to 1 (0 corresponding to the most brittle material and 1 the most ductile one). Compared to non-irradiated samples, plasticity index at 1 mN applied load reduced by 51% and 2% for PI and TiNT/PI, respectively after 3 h UV exposure indicating that TiNT protected the PI underneath from further damage.

**Acknowledgements** This work has been financially supported by the Indonesian Endowment Fund for Education (LPDP).

## Affiliations

**Christian Harito[1,2,3] · Dmitry V. Bavykin[1] · Brian Yuliarto[2,3] · Hermawan K. Dipojono[3] · Frank C. Walsh[1]**

1. Energy Technology Research Group, Faculty of Engineering and the Physical Sciences, University of Southampton, Southampton SO17 1BJ, UK

2. Advanced Functional Materials (AFM) Laboratory, Engineering Physics, Institut Teknologi Bandung, Bandung 40132, Indonesia

3. Research Centre for Nanosciences and Nanotechnology (RCNN), Institut Teknologi Bandung, Bandung 40132, Indonesia